\documentclass[aps,prr,twocolumn,superscriptaddress,showpacs,floatfix,nofootinbib,llncs]{revtex4-1}
\usepackage{graphicx}
\usepackage{amssymb}
\usepackage{amsmath}
\usepackage{color}
\usepackage{psfrag}
\usepackage{epsfig}
\usepackage{bbm}
\usepackage{bm}
\usepackage{hyperref}
\usepackage[normalem]{ulem}
\usepackage{amsthm}
\usepackage{epstopdf}
\usepackage{verbatim}
\usepackage{subfigure}
\usepackage[export]{adjustbox}
\usepackage{bbold}

\usepackage{array}
\usepackage{mathtools}

\DeclareMathOperator{\tr}{tr}

\newcolumntype{M}[1]{>{\centering\arraybackslash}m{#1}}

\begin{document}
\title{Space-time dual quantum Zeno effect: Interferometric engineering of open quantum system dynamics
}
\author{Jhen-Dong Lin}
\email{jhendonglin@gmail.com}
\affiliation{Department of Physics, National Cheng Kung University, 701 Tainan, Taiwan}
\affiliation{Center for Quantum Frontiers of Research \& Technology, NCKU, 70101 Tainan, Taiwan}

\author{Ching-Yu Huang}
\affiliation{Department of Physics, National Cheng Kung University, 701 Tainan, Taiwan}
\affiliation{Center for Quantum Frontiers of Research \& Technology, NCKU, 70101 Tainan, Taiwan}

\author{Neill Lambert}
\affiliation{Theoretical Quantum Physics Laboratory, RIKEN Cluster for Pioneering Research, Wako-shi, Saitama 351-0198, Japan}

\author{Guang-Yin Chen}
\affiliation{Department of Physics, National Chung Hsing University, Taichung 402, Taiwan}

\author{Franco Nori}
\affiliation{Theoretical Quantum Physics Laboratory, RIKEN Cluster for Pioneering Research, Wako-shi, Saitama 351-0198, Japan}
\affiliation{RIKEN Center for Quantum Computing (RQC), Wakoshi, Saitama 351-0198, Japan}
\affiliation{Department of Physics, The University of Michigan, Ann Arbor, 48109-1040 Michigan, USA}

\author{Yueh-Nan Chen}
\email{yuehnan@mail.ncku.edu.tw}
\affiliation{Department of Physics, National Cheng Kung University, 701 Tainan, Taiwan}
\affiliation{Center for Quantum Frontiers of Research \& Technology, NCKU, 70101 Tainan, Taiwan}

\date{\today}

\begin{abstract}
Superposition of trajectories, which modify quantum evolutions by superposing paths through interferometry, has been utilized to enhance various quantum communication tasks. However, little is known about its impact from the viewpoint of open quantum systems. Thus, we examine this subject from the perspective of system-environment interactions. We show that the superposition of multiple trajectories can result in quantum state freezing, suggesting a space-time dual to the quantum Zeno effect. Moreover, non-trivial Dicke-like super(sub)radiance can be triggered without utilizing multi-atom correlations.

\end{abstract}

\maketitle

\section{Introduction}
Controlling quantum dynamics is an essential part of quantum information science, which becomes richer and more challenging when considering dissipative open systems. In this regime, several unique control, or engineering, approaches are available. For example: (i) quantum reservoir engineering, which steers open system dynamics by directly manipulating an artificial environment~\cite{myatt2000decoherence,liu2011experimental,chiuri2012linear,PhysRevA.88.063806,PhysRevA.95.033610,liu2018experimental,PhysRevA.99.022107,garcia2020ibm,PhysRevLett.124.210502}; (ii) feedback-based control~\cite{PhysRevA.49.2133,PhysRevA.62.012105,PhysRevA.65.010101,wiseman2009quantum,zhang2017quantum,PhysRevA.81.062306}, where the dynamics is modified by closed-loop controls; (iii) dynamical decoupling~\cite{PhysRevLett.82.2417,PhysRevLett.83.4888,PhysRevLett.95.180501,yang2011preserving,du2009preserving,PhysRevLett.90.037901,de2010universal,liu2013noise,alonso2016generation}, which is an open-loop design to counter the effect of system-environment couplings; and (iv) the quantum Zeno effect~\cite{misra1977zeno,PhysRevA.41.2295,kofman2000acceleration,PhysRevLett.87.040402,PhysRevLett.89.080401,koshino2005quantum,PhysRevA.90.012101,chaudhry2016general,PhysRevA.77.062339,PhysRevA.80.062109,PhysRevA.82.022119,CAO2012349,ZHANG20131837,Ai2013}, where dissipation can be suppressed through frequent measurements on the open system.  

\begin{figure*}
\includegraphics[width = 1\linewidth]{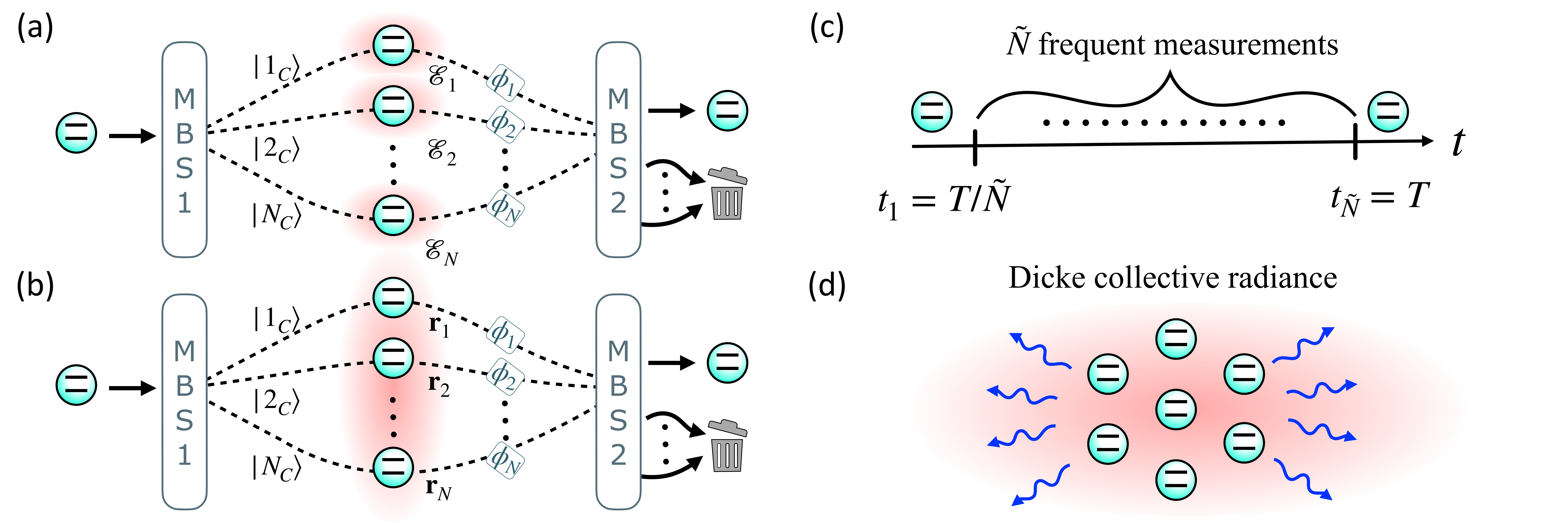}
\caption{Superposed quantum dynamics achieved by a multi-arm interferometer. To produce superposition of paths, the qubit is first sent into the multi-port beam splitter (MBS1) such that the path of the qubit is prepared in $|\chi_C\rangle = \sum_{i=1}^N |i_C\rangle /\sqrt{N}$. Thus, the qubit can travel through (a) different independent environments $\{\mathcal{E}_i\}$ or (b), different positions $\{\mathbf{r}_i\}$ of a single environment in a quantum superposition. One can further manipulate the interference effect between the evolution paths by using the other beam splitter (MBS2) along with phase shifters labeled by $\{\phi_i\}$. The modified dynamics can be obtained through a post-selection, as illustrated by the trash cans. For the independent-environments scenario, one can obtain a space-time dual to (c), usual quantum Zeno effect induced by a temporal sequence of measurements. For the indefinite-position scenario, one can obtain a collective decay akin to (d): the Dicke effect that can be observed by an ensemble of qubits embedded in a common environment.}\label{ill_sup_traj}
\end{figure*}

Recently, an interferometric scheme known as \textit{superposed trajectories} has drawn considerable research interest~\cite{PhysRevLett.91.067902,PhysRevA.72.012338,chiribella2019quantum,PhysRevA.101.012340,abbott2020communication,kristjansson2020resource,PhysRevResearch.3.013093,PhysRevD.103.065013,PhysRevD.102.085013,PhysRevLett.125.131602,ban2021two,ban2020relaxation,PhysRevA.103.032223}. This approach utilizes a quantum control of evolution paths to let the target system to go through different evolution paths in a quantum superposition. In principle, the superposition of paths can be implemented by a Mach-Zehnder type interferometer~\cite{PhysRevLett.73.58,PhysRevA.89.062316,RevModPhys.81.1051,nairz2003quantum,sadana2019double,carine2020multi,margalit2021realization}, as illustrated in Fig.~\ref{ill_sup_traj}. The quantum interference between different evolution paths can reduce the noise effect. Thus, it is beneficial for quantum communication~\cite{PhysRevLett.91.067902, PhysRevA.72.012338,chiribella2019quantum, PhysRevA.101.012340,abbott2020communication,kristjansson2020resource, PhysRevResearch.3.013093}, quantum metrology~\cite{lee2022steering} and quantum thermodynamics~\cite{chan2022maxwell}.  Also, it has potential applications in relativistic quantum theory~\cite{PhysRevD.103.065013, PhysRevD.102.085013,PhysRevLett.125.131602}. The ability to mitigate quantum decoherence has also stimulated the quantum open-system community to investigate the concept of superposed trajectories in more detail~\cite{ban2021two,ban2020relaxation, PhysRevA.103.032223}; however, various questions remain. For instance, in many previous works, only two paths were considered, which may limit the utility of the quantum interference effect. Further, these works only consider what we call the independent-environments scenario, where the environments inside the interferometer are considered to be separated and independent from each other. This simplified scenario may deviate from real-world considerations; for instance, these environments could either be correlated or be different regions of a single environment.

Here, we explore both open questions. First, we extend the exploration of the independent-environments scenario to multiple evolution paths.
Our primary result here is revealing an unexpected connection between superposed trajectories and the quantum Zeno effect. For concreteness, we consider the dissipative and the pure dephasing spin-boson models~\cite{breuer2002theory}. We find that quantum state freezing occurs when the number of superposed evolution paths reaches infinity. Moreover, we show that the effective decay can be, in general, characterized by the overlap-integral expression, which serves as a universal tool to study the quantum Zeno effect~\cite{kofman2000acceleration} and noise spectroscopy~\cite{bylander2011noise,PhysRevLett.107.230501,PhysRevLett.108.140403, PhysRevLett.129.030401}. 

Our result could also open a novel alternative approach to investigate the space-time “dual" quantum Zeno effect~\cite{PRXQuantum.2.040319,
PhysRevLett.126.060501}, because we replace the \textit{temporal sequence} of measurements performed at one location (of the atom) with a \textit{single} measurement done on the multiple paths followed by the atom in the interferometer. In other words, a temporal sequence of measurements in one location is replaced by a single measurement at one time, but over many paths (i.e., many locations).

Second, we consider an indefinite-position scenario, wherein an initially excited two-level atom is placed inside a bosonic vacuum where its position is indefinite because of the superposition of paths. The modified decay can exhibit signatures of both the superradiant and subradiant emission effects~\cite{PhysRev.93.99, PhysRevLett.76.2049,gross1982superradiance,chen2005proposal, PhysRevA.88.052320, PhysRevLett.90.166802,brandes2005coherent,chen2013examining,PhysRevA.98.063815}. It is well known that Dicke first proposed the idea of the superradiance effect induced by an ensemble of correlated atoms~\cite{PhysRev.93.99}, wherein the quantum correlations make them behave like a giant dipole moment. Our result implies that the formation of a giant dipole moment can be emulated by only one atom with superposed trajectories.We expect that this single-atom collective effect can be utilized to design new Dicke quantum batteries~\cite{PhysRevLett.120.117702,
quach2022superabsorption} and heat engines~\cite{PhysRevLett.128.180602}.

\section{Independent-environments scenario}
We formalize the scenario depicted in Fig.~\ref{ill_sup_traj}(a). We consider that the path of a traveling qubit $Q$ inside the interferometer is characterized by a quantum system $C$. More specifically, we introduce $N$ orthonormal states $\{|i_C\rangle\}_{i=1\cdots N}$ to describe $N$ possible paths for the qubit. When $C$ is prepared in the state $|i_C\rangle$, the qubit goes through the path labeled by $i$ and interacts with the environment $\mathcal{E}_i$. One can also prepare $C$ in a superposition state, i.e., a superposition of paths, by sending the qubit into a multi-port beam splitter [the MBS1 in Fig.~\ref{ill_sup_traj}(a)], such that $Q$ interacts with all environments $\left\{\mathcal{E}_i\right\}$ as a coherent superposition. Therefore, $C$ acts as a quantum control to determine which environment for $Q$ to interact with. Further, we assume that $C$ does not directly interact with these environments. Thus, inside the interferometer, the total Hamiltonian of $C$, $Q$, and $\left\{\mathcal{E}_i\right\}_{i=1\cdots N}$ can be written as 
\begin{equation}
H_{\text{tot}} = \sum_{i=1}^N |i_C\rangle\langle i_C| \otimes H_{Q\mathcal{E}_i} , \label{Htot}
\end{equation}
where $H_{Q\mathcal{E}_i}$ represents the interaction Hamiltonian of the qubit $Q$ and the environment $\mathcal{E}_i$.

The initial states of $Q$ and $\left\{\mathcal{E}_i\right\}$ are $\rho_{Q}(0)$ and $\left\{ \rho_{\mathcal{E}_i}(0)\right\}$, respectively. Also, by using MBS1, $C$ is prepared in 
\begin{equation}
    |\chi_C\rangle =\sum_{i=1}^N|i_C\rangle/\sqrt{N}.
\end{equation}
Thus, after passing through these environments, the reduced dynamics of $CQ$ complex is
\begin{align}
&\rho_{CQ}(t) = \frac{1}{N}\sum_{i,j=1}^N |i_C\rangle \langle j_C| \otimes \rho_{Q,i,j}(t)\nonumber\\
&\text{with}~~
\rho_{Q,i,j}(t)=\mathrm{tr}_{\{\mathcal{E}_k\}}\big[e^{-iH_{Q\mathcal{E}_i }t}~\rho_Q(0)\bigotimes_{l=1}^N \rho_{\mathcal{E}_l}(0)~e^{iH_{Q\mathcal{E}_j }t}\big].
\end{align}
Terms with $i=j$ describe the reduced dynamics obtained from sending the qubit into a single path with the label $i$, i.e., the single-path dynamics; while, the terms with $i\neq j$ (i.e., the off-diagonal terms) capture the quantum interference between the paths $i$ and $j$.

Before discarding $C$, one should perform another selective measurement on it to harness the quantum interference effect~\cite{PhysRevA.101.012340,abbott2020communication}. As shown in Fig.~\ref{ill_sup_traj}(a), this can be achieved by applying the second multi-port beam splitter (MBS2) with the phase shifters $\left\{\phi_i\right\}$ and selecting one of the output beams of the qubit. To illustrate this idea, we consider that the selective measurement is characterized by a projector
\begin{align}
    &P_{C,\bm{\phi}} = |\chi_{C,\bm{\phi}}\rangle \langle \chi_{C,\bm{\phi}}|\nonumber \\
    \text{with~~}&|\chi_{C,\bm{\phi}}\rangle =\sum_{i=1}^N \exp(i\phi_i)|i_C\rangle/\sqrt{N}.
\end{align}
The post-measurement state of $Q$ then reads 
\begin{align}
&\tilde{\rho}_{Q,\bm{\phi}}(t) =\langle \chi_{C,\bm{\phi}}|\rho_{CQ}(t)|\chi_{C,\bm{\phi}}\rangle \nonumber\\
&=\frac{1}{N^2}\sum_{i,j}e^{-i(\phi_i-\phi_j)}\rho_{Q,i,j}(t) \nonumber\\
&=\frac{1}{N}\rho_{Q,\mathrm{avg}}(t) +\frac{1}{N^2}\sum_{i\neq j}\left[e^{-i(\phi_i-\phi_j)}\rho_{Q,i,j}(t) \right].\label{rho_Q_t_unnorm}
\end{align}
Here, $\rho_{Q,\text{avg}}(t)=\sum_i \rho_{Q,i,i}(t)/N$ denotes the incoherent mixture of the single-path dynamics~\cite{megier2017eternal,PhysRevA.94.022118,breuer2018mixing,PhysRevA.101.062304,PhysRevA.103.022605}. Note that the normalized post-measured state is written as 
\begin{equation}
    \rho_{Q,{\bm{\phi}}}(t)=\tilde{\rho}_{Q,{\bm{\phi}}}(t)/\mathrm{tr}\left[\tilde{\rho}_{Q,{\bm{\phi}}}(t)\right].
\end{equation}
Equation~\eqref{rho_Q_t_unnorm} suggests that the interferometry-modification of the qubit dynamics originates from both the incoherent mixing of single-path dynamics and the interference effects, i.e., the off diagonal terms $\rho_{Q,i,j}(t)$ with $i\neq j$ and the phase shifts $\{\phi_i\} $.

To simplify the following discussions, we introduce two additional assumptions. First, we assume that all single-path dynamics are identical; that is, $\rho_Q(t)=\rho_Q^{\text{avg}}(t)=\rho_{Q,i,i}(t)~\forall i$ and $\beta(t) = \rho_{Q,i,j}(t)~\forall i\neq j$. In this case, the incoherent mixing does not yield a new dynamical process; thus, the modification is totally determined by the interference effect. This assumption holds when all of the environments are prepared in the same state and all of the qubit-environment interactions are identical. Second, we consider that each $\phi_i$ is either $0$ or $\pi$; and, hence, the projector associated with the selective measurement can be written as $P_{C,\bm{\phi}_n} = |\chi_{C,\bm{\phi}_n}\rangle\langle \chi_{C,\bm{\phi}_n}| $ with $n$ being the number of phase shifts that take the value $\pi$. The post-measurement unnormalized state of $Q$ can then be simplified as 
\begin{align}
    &\tilde{\rho}_{Q,{\bm{\phi}_n}}(t)=\rho_Q(t)/N+\left(R_{N,n}-1/N\right)\beta(t) \nonumber \\
    \text{with~~} &R_{N,n} = \left(N-n/N\right)^2.
\end{align}
One can find that $\rho_{Q,\bm{\phi}_{N/2-k}}(t)=\rho_{Q,\bm{\phi}_{N/2+k}}(t)$ for $k=0,1, \cdots, N/2-1$. Note that at $t=0$ we have
\begin{equation}
\rho_{Q,\bm{\phi}_n}(0)=
	\begin{cases}
	\rho_Q(0) &\text{for}~n\neq \frac{N}{2}\\
	\mathbf{0}&\text{for}~n=\frac{N}{2}
	\end{cases}.
\end{equation} 
When $n=N/2$, one obtains a null result because of the completely destructive interference, i.e., $\langle \chi_{\bm{\phi}_0}|\chi_{\bm{\phi}_{N/2}}\rangle=0 $, and thus, we naturally exclude this scenario from the rest of the discussions.

\section{Space-time dual quantum Zeno effect for the dissipative and the pure dephasing models}
We now focus on two different types of spin-boson interactions, the dissipative and the pure dephasing models. Without loss of generality, we work within the interaction picture; the interaction Hamiltonians are
\begin{align}
H^{\text{diss}}_{Q\mathcal{E}_i}(t) &= \sum_k g_k e^{i(\omega_q-\omega_k)t}\sigma_+ a_{i,k} + g_k^* e^{-i(\omega_q-\omega_k)} \sigma_-a^{\dagger}_{i,k}, \nonumber\\
H^{\text{deph}}_{Q\mathcal{E}_i}(t) &= \sigma_z \sum_k (g_k e^{-i\omega_k t} a_{i,k}+ g_k^* e^{i\omega_k t} a_{i,k}^{\dagger}).
\label{interaction_diss}
\end{align}
Here, $\omega_q$ denotes the energy gap between the excited state $|e\rangle$ and the ground state $|g\rangle$ of the qubit, $\sigma_z = |e\rangle \langle e|-|g\rangle \langle g|$, $\sigma_+=|e\rangle \langle g|$ ($\sigma_-=|g\rangle \langle e|$) represents the qubit raising (lowering) operator, and $a_{i,k}$ ($a_{i,k}^{\dagger}$) stands for the annihilation (creation) of the mode $k$ for the environment $\mathcal{E}_i$. Let us assume that the initial state of the total system is
\begin{equation}
    \rho_{\text{tot}}(0)=|\chi_C\rangle\langle\chi_C| \otimes \rho_Q(0)\otimes |\text{vac}\rangle\langle \text{vac}|,
\end{equation}
where $|\text{vac}\rangle =\bigotimes_{i=1}^N|\text{vac}_i\rangle$, with $a_{i,k} |\text{vac}_i\rangle =0$. The associated reduced dynamics can be obtained analytically; detailed derivations can be found in Appendix~\ref{derivations}. As pointed out in Refs.~\cite{ban2021two,ban2020relaxation, PhysRevA.103.032223}, superposed trajectories can modulate the quantum non-Markovian effect. Tunability also holds in our models; we discuss this in detail in Appendix~\ref{non-markovian}.

\begin{figure}
\includegraphics[width=1\linewidth]{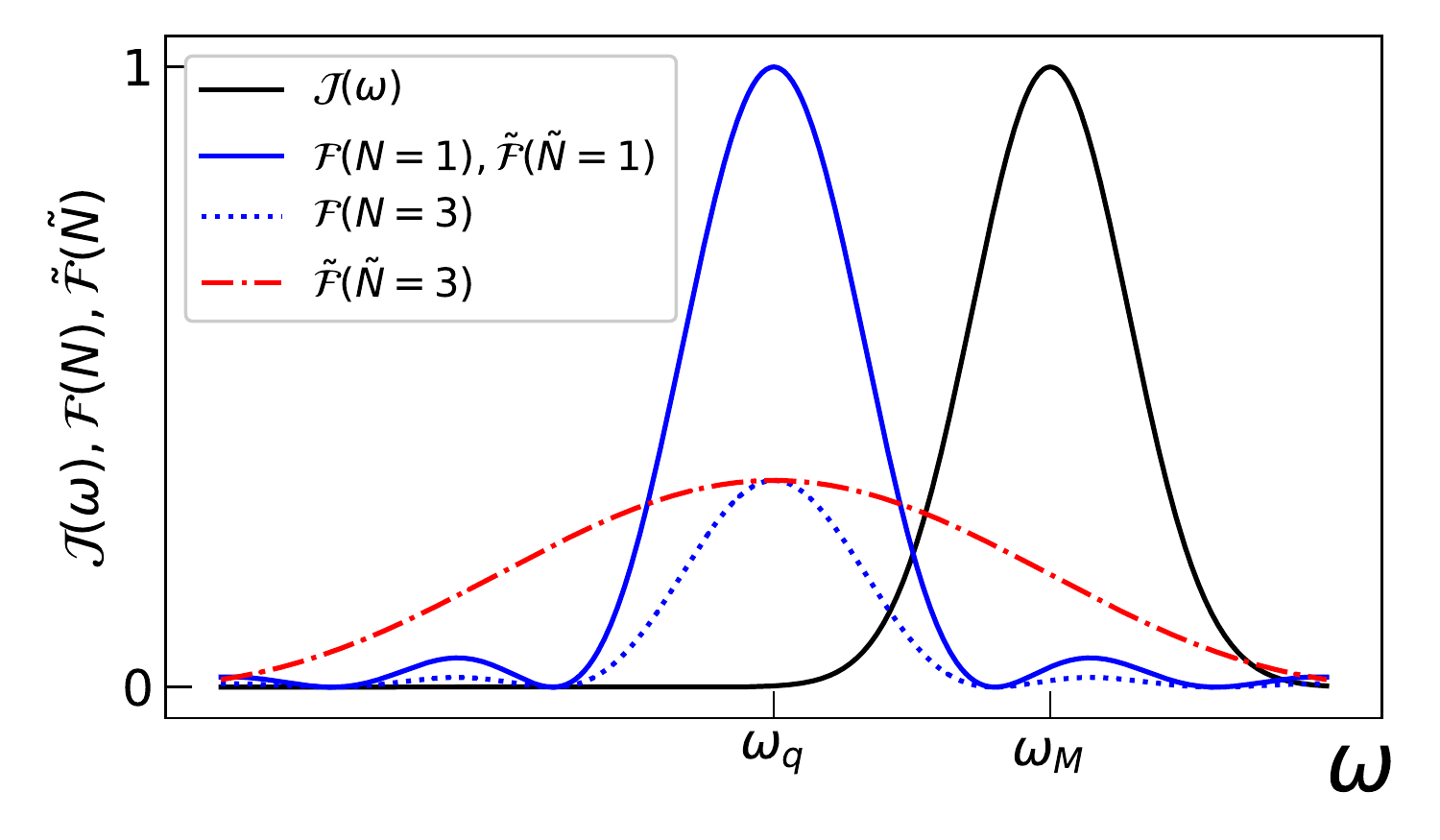}
\caption{
Effective decay within a given time $ t= \omega_q/5$ of the dissipative model for superposed trajectories and the usual quantum Zeno effect, which are described by a spectral density and the filter functions. $\mathcal{F}^{\text{diss}}(N)$, given by Eq.~\eqref{filter_superposed}, is the filter function with $N$ superposed trajectories with $n=0$. $\tilde{\mathcal{F}}(\tilde{N})$ is the filter function for the Zeno effect induced by $\tilde{N}$ periodic measurements of the qubit energy; it can be obtained by replacing $N$ with $\tilde{N}$ and $\mathrm{sinc}^2\big[(\omega-\omega_q)t/2\big]$ with $\mathrm{sinc}^2\big[(\omega-\omega_q)t/2\tilde{N}\big]$ in $\mathcal{F}(N)$. $\mathcal{J}(\omega)=\exp[-(\omega-\omega_M)^2/\Delta]$ represents a given spectral density with $\omega_M = 3\omega_q/2$ and $\Delta=\omega_q/5$. The qualitative difference between the traditional Zeno effect and the space-time dual Zeno effect induced by superposed trajectories is that $\mathcal{\tilde{F}}({\tilde{N}})$ smears out, whereas $\mathcal{F}(N)$ remains localized when increasing $\tilde{N}$ and $N$, respectively.
\label{ill_zeno}}
\end{figure}

\begin{figure}
\includegraphics[width=1\linewidth]{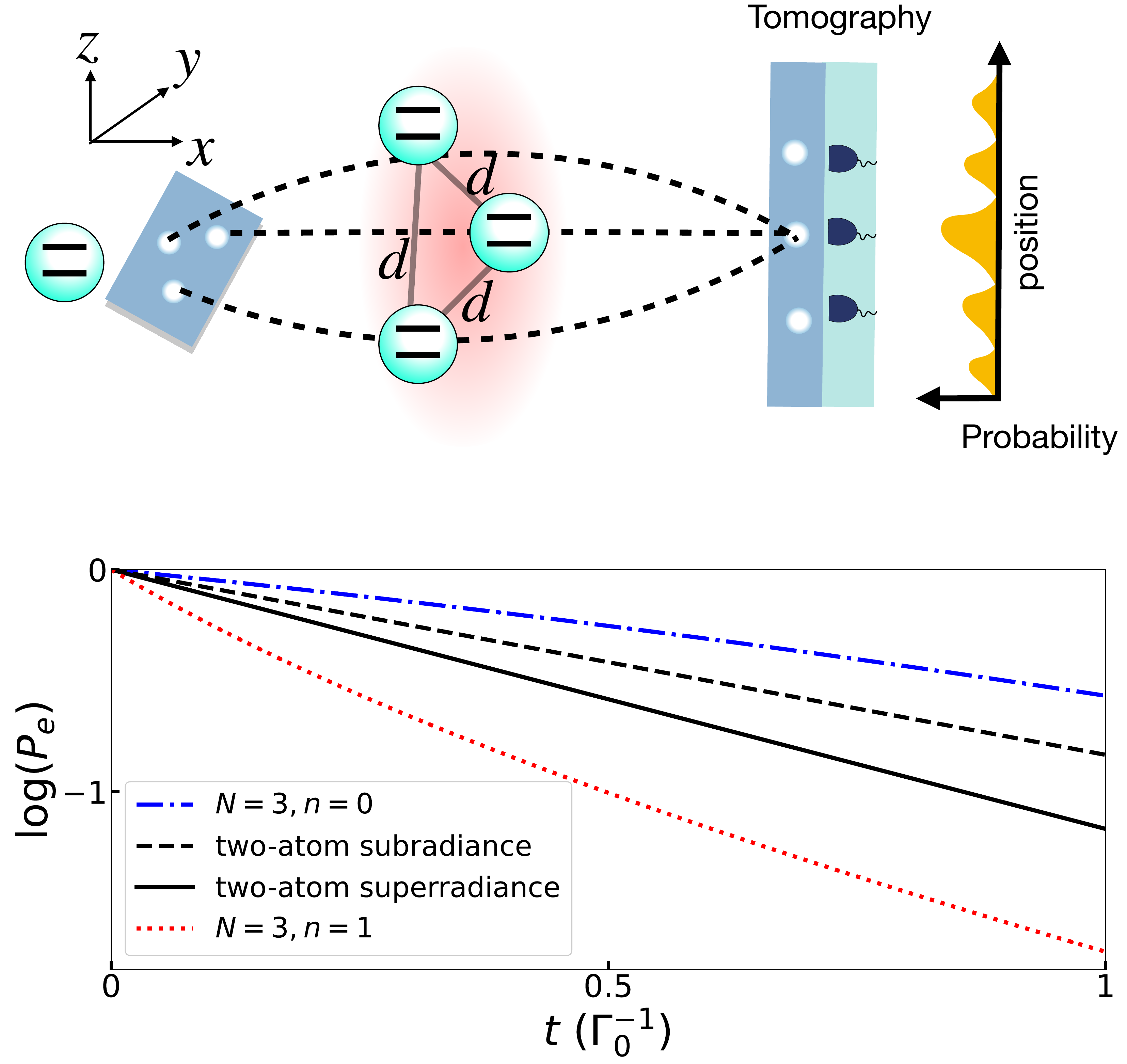}
\caption{Comparison of the two-atom super(sub)radiance, with qubits' distance $d$, with the single-atom decay modified by three superposed trajectories, where the position vectors form an equilateral triangle with edge length $d$. This can be implemented by a Young's type triple-slit experiment. One can place the atom detectors at certain positions and perform quantum state tomography to verify the super(sub) radiant dynamics. Here, we set the collective factor $\mathrm{sinc}(qd)=1/6$ and the spontaneous emission rate $\Gamma_0/\omega_q=0.01$. 
\label{figsup}}
\end{figure}

If we now prepare the qubit state in $|\psi^{\text{diss}}_Q(0)\rangle=|e\rangle $ and $|\psi^{\text{deph}}_Q(0)\rangle=(|e\rangle +|g\rangle )/\sqrt{2}$ for the dissipative and the pure dephasing models, respectively, then
\begin{align}
\lim_{N\rightarrow\infty} \rho^{\text{diss}}_{Q,\bm{\phi}_n}(t) &= |\psi_Q^{\text{diss}}(0)\rangle \langle \psi_Q^{\text{diss}}(0)|, \nonumber \\
\lim_{N\rightarrow\infty} \rho^{\text{deph}}_{Q,\bm{\phi}_n}(t) &= |\psi_Q^{\text{deph}}(0)\rangle \langle \psi_Q^{\text{deph}}(0)|~~\forall t
\end{align}
with $n$ being a finite positive integer. That is, the quantum states of the qubit are frozen when the number of paths goes to infinity.

To gain a deeper insight, we introduce the survival probability defined as 
\begin{equation}
p(t) = \tr [|\psi_Q(0)\rangle \langle \psi_Q(0)|~ \rho_{Q,\bm{\phi}_n}(t)].
\end{equation}
We also consider the decay factor $\gamma(t)$ associated with the survival probability $p(t) = \exp[-\gamma(t)]$, or equivalently, 
\begin{equation}
    \gamma(t) = -\log p(t).
\end{equation}
Within leading order in perturbation, the decay factor can be described by an overlap integral in a similar manner with the traditional quantum Zeno effect~\cite{kofman2000acceleration,PhysRevLett.87.040402,PhysRevLett.89.080401,koshino2005quantum,PhysRevA.90.012101,chaudhry2016general}; namely,
\begin{align}
\gamma(t) = \int d\omega~ \mathcal{J}(\omega) \mathcal{F}(\omega,t,N,n). \label{decay_factor}
\end{align}
Here, $\mathcal{J}(\omega)$ denotes the system-environment coupling spectral density, and the filter function $\mathcal{F}(\omega,t,N,n)$ can be expressed as
\begin{align}
\mathcal{F}^{\text{diss}}(\omega,t,N,n)&= \frac{N}{(N-2n)^2}t^2\mathrm{sinc}^2\left[\frac{(\omega-\omega_q)}{2}t\right]\nonumber\\
\mathcal{F}^{\text{deph}}(\omega,t,N,n) &= \frac{1}{2} \frac{N}{(N-2n)^2}\frac{1-\cos(\omega t)}{\omega^2}. \label{filter_superposed}
\end{align}
Note that $\mathrm{sinc}(x)=\sin(x)/x$. Equations \eqref{decay_factor} and \eqref{filter_superposed} show that one can modify the decay by either introducing different number of paths $N$ or modulating the phase shifts, i.e., changing the value $n$. 

We emphasize that this overlap-integral expression can be  derived for a more general class of open-system models without imposing the two assumptions mentioned earlier, namely, (1) all environments are identical, and (2) each $\phi_i$ is either $0$ or $\pi$ (see Appendix~\ref{general_zeno} for detailed derivations). The only requirement for the validity of such an expression is the weak system-environment couplings, such that the reduced dynamics can be perturbatively approximated.

One distinct feature of the traditional Zeno effect is that frequent measurements broaden the filter functions, as shown in Fig.~\ref{ill_zeno}. For dissipative processes, the broadening has been interpreted as a consequence of energy-time uncertainty~\cite{kofman2000acceleration} because the system energy is measured frequently. Superposed trajectories only modify the overall magnitude of the filter function without broadening. This is physically reasonable because the open system is not measured frequently in the superposed trajectories scenario, and therefore, the energy-time uncertainty does not occur.

\section{Indefinite-position scenario and single-atom Dicke-like decay} We now consider a scenario that generates behavior normally observed under the Dicke effect. The simplest model to illustrate the Dicke effect is that of two identical two-level atoms, with an energy gap $\omega_q$, embedded in a bosonic vacuum (see, e.g., ~\cite{brandes2005coherent,PhysRevA.98.063815} for instance). By utilizing either the Fermi--Gorden rule or the master equation approach, one can predict two split decay rates, $\Gamma_\pm=\Gamma_0[1\pm \mathrm{sinc}(qd)]$, where $\Gamma_0$ represents the single-atom spontaneous decay rate, $d$ is the distance between two atoms and $q=\omega_q/c$, with $c$ being the speed of light in vacuum. The factor $\mathrm{sinc}(qd)$ can be interpreted as the collective effect for this two-atom model. Here, $\Gamma_+>\Gamma_0$ ($\Gamma_-<\Gamma_0)$, which is known as Dicke superradiance (subradiance), if $\mathrm{sinc}(qd)>0$.

Inspired by this model, we consider that a single qubit $Q$ interacts with a single bosonic vacuum, where the location of $Q$ is coherently controlled by $C$, as depicted by Fig.~\ref{ill_sup_traj}(b). We model the total Hamiltonian as 
\begin{align}
    &\tilde{H}_{\text{tot}}=\sum_{i=1}^N |i_C\rangle\langle i_C|\otimes \tilde{H}(\mathbf{r}_i) \nonumber\\
    \text{with~~}&\tilde{H}(\mathbf{r}_i) = \omega_q\sigma_z/2 + \sum_{\mathbf{k}} \omega_k a^\dagger_\mathbf{k}a_\mathbf{k}\nonumber\\
    &~~~~~~~~~+\sigma_x \sum_{\mathbf{k}} g_{\mathbf{k}} \left( a_\mathbf{k}e^{i\mathbf{k}\cdot \mathbf{r}_i}+ a^\dagger_\mathbf{k}e^{-i\mathbf{k}\cdot \mathbf{r}_i} \right).
\end{align}
Here, $\{\mathbf{r}_i\}$ denotes the possible positions of $Q$ that are controlled by $C$. By taking Born--Markov and secular approximations, the time evolution of the $CQ$ complex is governed by the following master equation
\begin{align}
\frac{\partial\rho_{CQ}(t)}{\partial t}=&\Gamma_0 \sum_{i=1}^N L_i \rho_{CQ}(t) L_i^{\dagger} -\frac{1}{2}\{L_i^\dagger L_i,\rho_{CQ}(t)\} \nonumber \\
+&\Gamma_0 \sum_{i\neq j } \mathrm{sinc}\big(q|\mathbf{r}_i-\mathbf{r}_j| \big) L_i \rho_{CQ}(t) L_j^\dagger. \label{master_eq_dicke}
\end{align}
Here, $L_i = |i\rangle \langle i| \otimes \sigma_-$. The evolution governed by the Lamb-shifted Hamiltonian is neglected for simplicity. In Eq.~\eqref{master_eq_dicke}, we observe the emergence of the factor $\mathrm{sinc}\big(q|\mathbf{r}_i-\mathbf{r}_j| \big)$, which is present in the above two-atom example, and this suggests that a Dicke-like collective effect also plays a non-trivial role. 

We initialize the state of systems $C$ and $Q$ as $\rho_{CQ}(0) = |\chi_C\rangle \langle \chi_C| \otimes |e\rangle \langle e|$ to investigate the effective population decay of $Q$ modified by the superposed trajectories. For simplicity, we consider $|\mathbf{r}_i-\mathbf{r}_j|=d,~\forall i\neq j$. That is, the position vectors $\{\mathbf{r}_i\}$ form an equilateral triangle for $N=3$ or a regular tetrahedron for $N=4$ with the edge length $d$. Following the procedure described in the previous sections, we perform the projective measurement $P_{\chi_{\bm{\phi}_n}}$ on $C$ so that the effective dynamics of the excited state population after the post-selection can be expressed as
\begin{align}
P_e(t,N,n) = \frac{\langle \chi_{\bm{\phi}_n}|\langle e|\rho_{CQ}(t)|\chi_{\bm{\phi}_n}\rangle|e\rangle}{\tr\big[\langle \chi_{\bm{\phi}_n}|\langle e|\rho_{CQ}(t)|\chi_{\bm{\phi}_n}\rangle|e\rangle\big]} \nonumber ~~~~~~~~~~\\
=\frac{R_{N,n} e^{-\Gamma_0 t}}{\frac{1}{N}+\big(R_{N,n}-\frac{1}{N}\big)\big[e^{-\Gamma_0 t}+\mathrm{sinc}(qd)(1-e^{-\Gamma_0 t})\big]}. 
\end{align}
Therefore, the effective decay of $Q$ depends on the factors $(N,n)$, which are determined by the superposed trajectories setup, and most importantly, the collective factor $\mathrm{sinc}(qd)$. In Fig.~\ref{figsup}, we present a comparison between the two-atom super(sub)radiant decay with distance $d$ such that $\mathrm{sinc}(qd)$, and a single-atom effective decay from three superposed trajectories. 
We consider that the position vectors form an equilateral triangle with edge length $d$. We set the spontaneous emission rate $\Gamma_0/\omega_q$ as $0.01$. The single-atom effective decay can either be greater than the two-qubit superradiance, i.e., $(N=3,n=1)$, or less than the two-atom subradiance, i.e., $(N=3,n=0)$. 
 
For the experimental realization, a Young's experiment (as illustrated in Fig.~\ref{figsup}), which has been applied to large molecules~\cite{nairz2003quantum}, can be considered as a natural test-bed for such superposed trajectories. A beam of atoms passes through a plate pierced by two or three slits, and therefore, the atom may interact with the environment at different locations. Atom detectors can be placed at different positions (on the right-hand side) to verify the super(sub) radiant effective dynamics. This is equivalent to performing a measurement on the path degrees of freedom and selecting the associated atoms. The modified dynamics can then be obtained by performing quantum state tomography on the selected atoms.
 
Although the effective decay can be modified similarly with the traditional Dicke effect, there exist non-trivial differences between these two distinct results. First, quantum correlations between multiple atoms, which is the most important ingredient for the traditional Dicke effect, are not present in the superposed trajectories because there is only one atom in the system. Second, it is known that the strongest superradiant effect for the traditional Dicke effect occurs in the so-called small sample limit, i.e., $q|\mathbf{r}_i - \mathbf{r}_j|\ll 1,~\forall~ i,j$. However, this is \textit{not} the case for superposed trajectories. The equation for $\tilde{H}_{\text{tot}}$ indicates that the superposition of paths cannot create the indefiniteness of the qubit position when all position vectors are identical, i.e., $|\mathbf{r}_i - \mathbf{r}_j|=0,~\forall~ i,j$; therefore, $\lim_{\{q|\mathbf{r}_i - \mathbf{r}_j|\rightarrow 0\}_{i,j}}P_e(t,N,n)=\exp(-\Gamma_0 t ),~\forall~ N,n$.

\section{Summary and outlook}
We studied the effects of superposed trajectories from the perspective of open quantum systems. We demonstrate a space-time dual Zeno effect when introducing multiple superposed trajectories for independent-environments scenario. More specifically, we find that it is possible to express the effective decay in terms of an overlap integral. This result provides a novel physical intuition to the problem, and we expect that it could be applicable to dynamical control~\cite{PhysRevLett.87.270405} and noise spectroscopy~\cite{bylander2011noise,PhysRevLett.107.230501,PhysRevLett.108.140403,
PhysRevLett.129.030401}, which is also based on overlap integrals. Moreover, it would be interesting to investigate whether the proposed interferometric setup can trigger a space-time dual of measurement-induced phase transitions~\cite{PRXQuantum.2.040319,
PhysRevLett.126.060501}, which is an application of the quantum Zeno effect to many-body physics. We leave this as a promising future work.

We also considered an indefinite-position scenario and demonstrated that the Dicke-like superradiant (subradiant) decay, usually observed by an ensemble of atoms, can be generated by only one atom with multiple superposed trajectories. One can naturally ask whether it is possible to induce an effective superabsorption~\cite{higgins2014superabsorption,
yang2021realization}, which could then open new possibilities to design quantum batteries~\cite{PhysRevLett.120.117702,
quach2022superabsorption} or quantum heat engines~\cite{PhysRevLett.128.180602}.

On the other hand, there exists another type of quantum-controlled evolution that induces indefinite causal order~\cite{oreshkov2012quantum,rubino2017experimental, PhysRevLett.120.120502,zych2019bell} of quantum processes. The investigation of this approach from the perspective of open systems is still an open question.

\section{Acknowledgement} This work is supported partially by the Ministry of Science and Technology, Taiwan, Grants No. MOST 111-2123-M-006-001 and the Army Research Office (under Grant No. W911NF-19-1-0081). G.Y.C. is partially supported by the National Center for Theoretical Sciences and the Ministry of Science and Technology, Taiwan [Grant No. MOST
110-2123-M-006-001. and MOST 110-2112-M-005-002]. F.N. is supported in part by: Nippon Telegraph and Telephone Corporation (NTT) Research, the Japan Science and Technology Agency (JST) [via the Quantum Leap Flagship Program (Q-LEAP) program, and the Moonshot R\&D Grant Number JPMJMS2061], the Japan Society for the Promotion of Science (JSPS) [via the Grants-in-Aid for Scientific Research (KAKENHI) Grant No. JP20H00134], the Army Research Office (ARO) (Grant No. W911NF-18-1-0358), the Asian Office of Aerospace Research and Development (AOARD) (via Grant No. FA2386-20-1-4069). F.N. and N.L. acknowledge support from the Foundational Questions Institute Fund (FQXi) via Grant No. FQXi-IAF19-06. N.L. is partially supported by JST PRESTO through Grant No. JPMJPR18GC.

\appendix

\section{Derivations of the spin-boson models \label{derivations}}
\subsection{The dissipative model}
The Hamiltonian of the dissipative spin-boson model is described by
\begin{equation}
H^{\text{diss}}_{Q\mathcal{E}_i}(t) = \sum_k g_k e^{i(\omega_q-\omega_k)t}\sigma_+ a_{i,k} + g_k^* e^{-i(\omega_q-\omega_k)} \sigma_-a^{\dagger}_{i,k}.
\end{equation}
Here, $\omega_q$ denotes the energy gap between the excited state $|e\rangle$ and the ground state $|g\rangle$ for the qubit $Q$, $\sigma_+=|e\rangle \langle g|$ ($\sigma_-=|g\rangle \langle e|$) represents the qubit raising (lowering) operator, and $a_{i,k}$ ($a_{i,k}^{\dagger}$) stands for the annihilation (creation) of the mode $k$ for the environment $\mathcal{E}_i$.

This model can be solved analytically in the single-excitation subspace spanned by the following basis states:
\begin{align}
\Big\{ &|\psi_{i,g}\rangle = |i_C\rangle \otimes |g\rangle \otimes |\text{vac}\rangle,~|\psi_{i,e}\rangle =|i_C\rangle \otimes |e\rangle\otimes |\text{vac}\rangle \nonumber \\
&|\psi_{i,k_j}\rangle =|i_C\rangle \otimes |g\rangle \otimes |k_j\rangle \Big\}_{i,j=1,\cdots,N}, 
\end{align}
where $|\text{vac}\rangle =\bigotimes_{i=1}^N|\text{vac}_i\rangle$ with $a_{i,k} |\text{vac}_i\rangle =0$ and $|k_j\rangle =a_{j,k}^{\dagger}|\text{vac}\rangle$. Whenever the initial state of the total system is expanded by the aforementioned bases, the quantum state at time $t$ can be written as
\begin{align}
|\Psi^{\text{diss}}_\text{tot}(t)\rangle =&\sum_i\Big[c_{i,g}(t)|\psi_{i,g}\rangle +c_{i,e}(t)|\psi_{i,e}\rangle \nonumber\\
	&~~~+\sum_{j,k}c_{i,k_j}(t)|\psi_{i,k_j}\rangle \Big]
\end{align} 
Further, the amplitudes satisfy the following coupled differential equations:
\begin{equation}
 \begin{cases}
 		\dot{c_{i,g}}(t)=0\\
 		\dot{c_{i,e}}(t) = -i\sum_k g_k e^{i(\omega_q-\omega_k)t }c_{i,k_i}(t)\\
 		\dot{c_{i,k_j}}(t)=-i\delta_{i,j }g_k^*e^{-i(\omega_q-\omega_k)t}c_{i,e}(t)
 \end{cases}~\text{for}~ i,j=1\cdots N. \label{coupled_eqs_diss}  
\end{equation}
Assuming that the environments are initially prepared in the vacuum state, i.e., $c_{i,k_j}(0)=0,~\forall i,j$, Eq.~\eqref{coupled_eqs_diss} can be analytically solved by Laplace transformation as

\begin{align}
&\begin{cases}
		&c_{i,g}(t) =c_{i,g}(0) \\
		&c_{i,e}(t) =c_{i,e}(0)G(t) \\
		&c_{i,k_j}(t) = -i\delta_{i,j} g_k^*\int_0^tdt^\prime e^{-i(\omega_q-\omega_k)t^\prime}c_{i,e}(t^\prime)
\end{cases} \nonumber\\
&\text{with}~G(t) = \mathcal{L}^{-1}\big[\frac{1}{s+\hat{f}(s)}\big] \nonumber \\
&\text{and}~\hat{f}(s) = \mathcal{L}\Big[\int_0^\infty d\omega~\mathcal{J}(\omega)e^{i(\omega_q-\omega)t} \Big].
\end{align}
Here, $G(t) = \mathcal{L}^{-1}\big[\frac{1}{s+\hat{f}(s)}]$ coincides with the dissipation function for the single-path dynamics, wherein $\mathcal{J}(\omega)=\sum_k |g_k|^2\delta(\omega-\omega_k)$ represents the spectral density function, and $\mathcal{L}$ and $\mathcal{L}^{-1}$ denote the Laplace and inverse Laplace transformations, respectively.
 
Assuming that the initial state of the total system is written as

\begin{align}
|\Psi^{\text{diss}}_\text{tot}(0)\rangle &=|\chi_C\rangle \otimes \big(c_g(0)|g\rangle +c_e(0)|e\rangle \big)\otimes |\text{vac}\rangle \nonumber \\
&=\sum_{i=1}^N\frac{1}{\sqrt{N}}\big(c_g(0)|\psi_{i,g}\rangle+c_e(0)|\psi_{i,e}\rangle \big), \label{solution_diss}
\end{align}
where $|\chi_C\rangle =\sum_{i=1}^N |i_C\rangle /\sqrt{N}$. The reduced dynamics of systems $C$ and $Q$ can then be written as
\begin{align}
\rho^{\text{diss}}_{CQ}&(t)\nonumber\\=&\frac{1}{N}\sum_{i,j}|i_C\rangle \langle j_C|\otimes \big(|c_g(0)|^2|g\rangle \langle g|+c_g(0)c_e(t)^*|g\rangle \langle e|\nonumber \\
&~~~~~~~~~~~~~~~~~~+c_e(t)c_g(0)^*|e\rangle \langle g|+|c_e(t)|^2|e\rangle \langle e|\big)\nonumber\\
&+\sum_i|i_C\rangle \langle i_C|\otimes |c_{i,k_i}(t)|^2|g\rangle \langle g|,
\end{align}
with $c_e(t)=c_e(0)G(t).$
Consequently, the unnormalized post-measurement state will be
\begin{align}
\tilde{\rho}_{Q,\bm{\phi}_n}^{\text{diss}}(t)=&\langle \chi_{C,\bm{\phi}_n}|\rho^{\text{diss}}_{CQ}(t)|\chi_{C,\bm{\phi}_n}\rangle\nonumber\\
=&\big[R_{N,n}|c_g(0)|^2 +\frac{1}{N}|c_e(0)|^2(1-|G(t)|^2)\big]|g\rangle \langle g|\nonumber
\\&+ R_{N,n} G^*(t)c_g(0)c_e^*(0)|g\rangle \langle e|\nonumber
\\&+R_{N,n} G(t) c_e(0)c_g(0)^*|e\rangle \langle g|\nonumber
\\&+R_{N,n} |G(t)|^2|c_e(0)|^2|e\rangle \langle e|
\label{unnorm_post_measured_state_diss}
\end{align} 
with $R_{N,n} =(N-2n)^2/N^2$. 

\subsection{The pure dephasing model}
Let us now consider the pure dephasing model, for which the interaction Hamiltonian is described as
\begin{equation}
H^{\text{deph}}_{Q\mathcal{E}_i}(t) = \sigma_z\otimes \sum_k (g_k e^{-i\omega_k t} a_{i,k}+ g_k^* e^{i\omega_k t} a_{i,k}^{\dagger}),
\label{interaction_deph}
\end{equation}
where $\sigma_{z} = |e\rangle \langle e| -|g\rangle \langle g|$. For this model, the unitary operator of the total system can be analytically derived as 
\begin{align}
U_{\text{tot}}^{\text{deph}}(t) &= \mathcal{T}_+\exp \left[-i\int_0^t H_{\text{tot}}^{\text{deph}}(t^\prime)dt^\prime\right]\nonumber\\
&=\sum_{i=1}^N |i\rangle \langle i|\otimes U^{\text{deph}}_{Q\mathcal{E}_i}(t).
\end{align}
Here,
\begin{align}
U^{\text{deph}}_{Q\mathcal{E}_i}(t)=\exp[if(t)]\prod_k[&|e\rangle\langle e|\otimes D_i(\alpha_k)+\nonumber\\&|g\rangle\langle g|\otimes D_i(-\alpha_k)], 
\end{align}
where $D_i(\alpha_k)=\exp(\alpha_k a_{i,k}^{\dagger} -\alpha_k^* a_{i,k})$ denotes the bosonic displacement operator with $\alpha_k = g_k^*(1-e^{i\omega_k t})/\omega_k$, and $f(t)=-\int_0^t dt^\prime_1 \int_0^{t^\prime_1}dt^\prime_2 \sum_k |g_k|^2 \sin[\omega_k(t^\prime_2-t^\prime_1)]$ characterizes an unimportant global phase. 

In the main text, we stated that the environments are prepared in the vacuum states. However, this model can be solved when the environments are prepared in thermal equilibrium states as well.  
Therefore, we now consider the initial state of the total system as
\begin{align}
&\rho^{\text{deph}}_\text{tot}(0)=|\chi_C\rangle\langle\chi_C| \otimes \rho_Q(0)\bigotimes_{i=1}^N \rho_{\mathcal{E}_i,T}, \nonumber\\
\text{with}~~&\rho_{\mathcal{E}_i,T} = \prod_k\big[1-e^{-\omega_k/k_B T}\big]e^{-\omega_k a^{\dagger}_{i,k} a_{i,k}/k_B T},
\end{align}
where $k_B$ and $T$ denote the Boltzmann constant and the temperature of the environments, respectively.  

After tracing out the environments, the reduced state of the systems $C$ and $Q$ can be obtained as  
\begin{align}
\rho^{\text{deph}}_{CQ}(t)=\frac{1}{N}\big[&\sum_{i=1}^N |i_C\rangle \langle i_C|\otimes \rho_Q(t)+\nonumber\\&\sum_{i\neq j}|i_C\rangle \langle j_C|\otimes \sqrt{\phi_T(t)}\rho_Q(0)\big],
\end{align}
where 
\begin{align}
\langle e|\rho_Q(t) |e\rangle &= \langle e|\rho_Q(0) |e\rangle, \nonumber \\  
\langle g|\rho_Q(t) |g\rangle &= \langle g|\rho_Q(0) |g\rangle, \nonumber \\ 
\langle e|\rho_Q(t) |g\rangle &= \langle e|\rho_Q(0) |g\rangle \phi_T(t)=(\langle g|\rho_Q(t) |e\rangle)^*\nonumber.\\
\end{align}
Here, $\phi_T(t) = 4\int_0^\infty d\omega\frac{\mathcal{J}(\omega)}{\omega^2}\coth(\omega/2k_BT)[1-\cos(\omega t)]$ represents the dephasing factor for the single-path dynamics.
The unnormalized post-measurement state can be written as 
\begin{align}
 \tilde{\rho}_{Q,\bm{\phi}_n}^{\text{deph}}(t)=&\langle \chi_{C,\bm{\phi}_n}|\rho^{\text{deph}}_{CQ}(t)|\chi_{C,\bm{\phi}_n}\rangle \nonumber\\
&=\frac{1}{N}\rho_Q(t)+\left(R_{N,n}-\frac{1}{N}\right)\sqrt{\phi_T(t)}\rho_Q(0)
\label{unnorm_post_measured_state_deph}.
\end{align} 
The normalized state retains a pure dephasing dynamics for the pure dephasing model with a dephasing function modified as
\begin{equation}
\Phi(t,N,n)=\frac{\phi_T(t)+[(N-1)-\frac{4n}{N}(N-n)\big]\sqrt{\phi_T(t)}}{1+[(N-1)-\frac{4n}{N}(N-n)\big]\sqrt{\phi_T(t)}}.
\end{equation}

\section{Full-time dynamics and quantum non-Markovian effects \label{non-markovian}}
\begin{figure}
\includegraphics[width=1\linewidth]{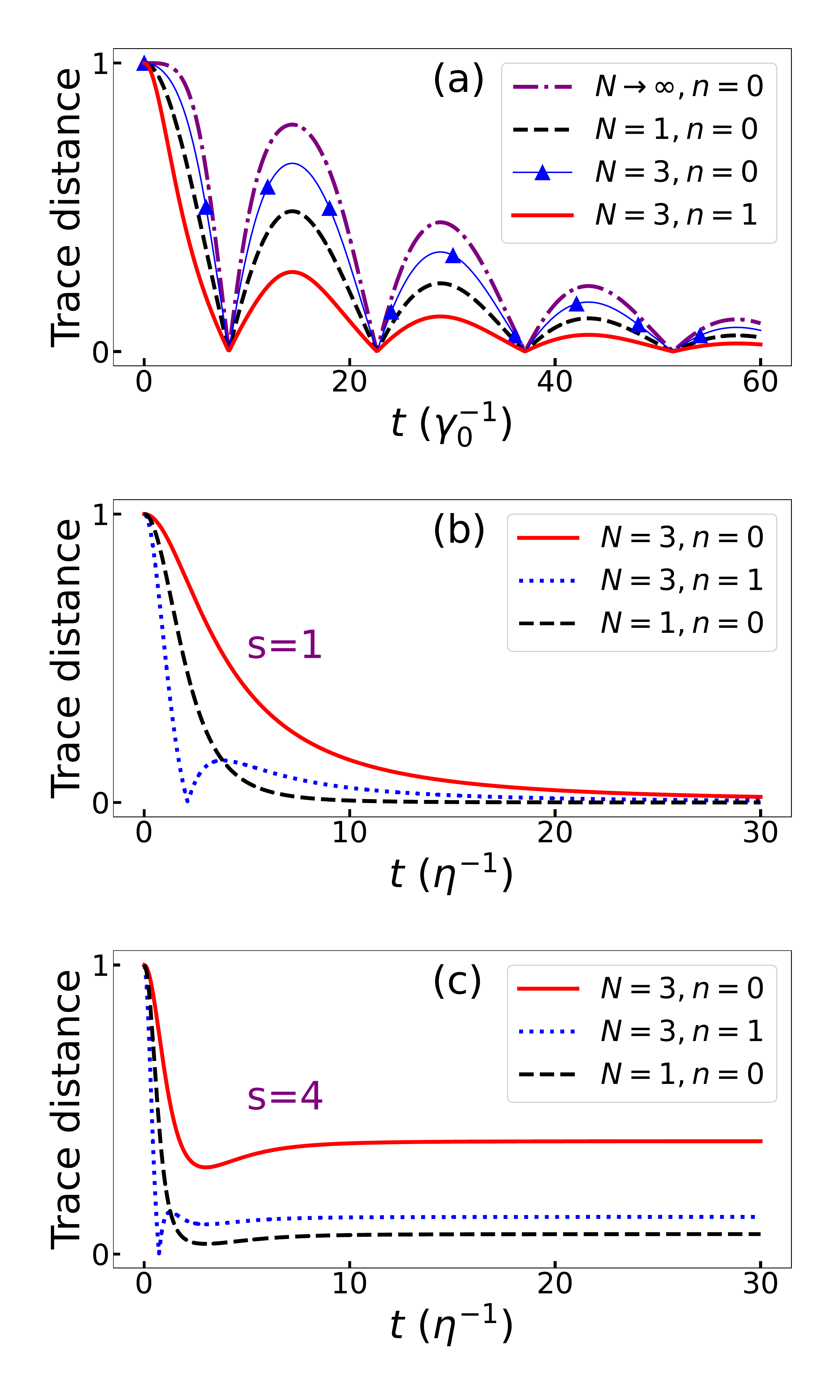}
\caption{Time evolutions of the trace distance for different factors $(N,n)$. (a) For the dissipative model, we consider the Lorentzian spectral density given by Eq.~\eqref{drude} with $\lambda/\gamma_0 = 0.1$. For the pure dephasing model, we consider a family of Ohmic spectral densities given by Eq.~\eqref{ohmic} with $\eta=1/3$ and $\omega_c=1$. (b) We present result for $s=1$ (Ohmic), where the single-path evolution experiences Markovian monotonic dephasing. (c) Further, we present result for $s=4$ (super-Ohmic), where non-monotonic behavior and coherence trapping can be observed for single-path dynamics.  
\label{non_markovian}}
\end{figure}
We now discuss the full-time dynamics and associated non-Markovian effects. A well-known indicator of quantum non-Markovianty is the non-monotonic behavior of the trace distance~\cite{PhysRevLett.103.210401, RevModPhys.88.021002}, which quantifies the distinguishability between quantum states. We consider an initial state pair $[\rho_{Q,+}(0),\rho_{Q,-}(0)]$, where
\begin{align}
\rho_{Q,+}(0) &= \frac{1}{2}(|e\rangle +|g\rangle )(\langle e| +\langle g|) \nonumber\\
\rho_{Q,-}(0) &= \frac{1}{2}(|e\rangle -|g\rangle )(\langle e| -\langle g|).
\end{align}
The time evolutions of the trace distances for the dissipative and pure dephasing models can be derived as
\begin{align}
\mathcal{D}&[\rho_{Q,\bm{\phi}_n,+}^{\text{diss}}(t),\rho_{Q,\bm{\phi}_n,-}^{\text{diss}}(t)] \nonumber \\
&= \frac{2(N-2n)^2|G(t)|^2}{[(N-2n)^2-N]|G(t)|^2 + (N-2n)^2 + N}\\
 \text{and}\nonumber\\
\mathcal{D}&[\rho_{Q,\bm{\phi}_n,+}^{\text{deph}}(t),\rho_{Q,\bm{\phi}_n,-}^{\text{deph}}(t)]=|\Phi(t,N,n)|,
\end{align}
where $\mathcal{D}(A,B)$ denotes the trace distance between $A$ and $B$. 

Taking the time derivative shows that the trace distances monotonically decrease whenever $|G(t)|^2$ and $|\Phi(t, N,n)|^2$ are also monotonic decreasing functions. For the dissipative model, the criterion of the monotonically decrease for the superposed trajectories coincides with that of the single-path dynamics (i.e., $d|G(t)|^2/dt \leq 0$). Therefore, non-monotonic behavior cannot be activated by superposed dynamics whenever the single-path dynamics are monotonic decreasing.

We present the numerical results for the non-Markovian dynamics in Fig.~\ref{non_markovian}. For the dissipative model, we consider the Lorentzian spectral density expressed as
\begin{equation}
\mathcal{J}_{\text{L}}(\omega)=\frac{1}{2\pi}\frac{\gamma_0 \lambda^2}{(\omega_q-\omega)^2+\lambda^2} \label{drude}
\end{equation}
with the width $\lambda$ and the coupling strength $\gamma_0$. The dynamics of the trace distance shows oscillating behavior when $\gamma_0>\lambda/2$; this criterion also holds for superposed trajectories. In Fig.~\ref{non_markovian}(a), we consider $\lambda = 0.1 \gamma_0$, for which oscillations can be observed. The magnitude of the oscillations, i.e., the strength of the non-Markovian effect, can either be enhanced or suppressed based on the factors $(N,n)$ in comparison with the single-path dynamics, i.e., $(N=1,n=0)$.

Let us now consider the pure dephasing model. Here, we consider a family of Ohmic spectral density parameterized as
\begin{equation}
\mathcal{J}_{\text{Ohmic}}(s,\omega) = \eta \omega^{s}\omega_c^{1-s}\exp\left(-\frac{\omega}{\omega_c}\right) \label{ohmic}
\end{equation}
with the coupling strength $\eta=1/3$, Ohmicity $s$, and the cut-off frequency $\omega_c=1$. We present the results for $s=1$ and $s=4$, which show single-path Markovian and non-Markovian dephasing, respectively. In the single-path Markovian regime $(s=1)$, we find that the interferometric engineering mitigates the dephasing process for $(N=3,n=0)$. Further, the non-monotonic behavior for $(N=3, n=1)$ implies that superposed trajectories can lead to non-Markovian dynamics even when single-path dynamics is Markovian (in contrast to the dissipative case). The trace distance (or equivalently the modified dephasing function) experiences a sudden death, and a sudden revival during the dephasing process. For the case $s=4$, the dephasing process is mitigated when $(N=3,n=0)$ and sudden death (revival) occurs when $(N=3,n=1)$. In addition, one can also find that the superposed trajectories can enhance another signature of non-Markovian effect known as coherence trapping~\cite{PhysRevA.89.024101}, where the coherence saturates to a finite value.

Quantum Markovianity is usually defined through the divisibility of dynamics characterized by a family of completely positive and trace-preserving (CPTP) maps. For the pure dephasing model, the concept of (non-)Markovianity can be applied directly because the effective dynamics of the post-measurement state remains a pure dephasing process that can be described by CPTP maps. However, for the dissipative model, the dynamics cannot be characterized by CPTP maps, because $\mathrm{tr}[\rho_{Q,\bm{\phi}_n}^{\text{diss}}(t)]$ is dependent on the initial state of the qubit $Q$. Nevertheless, the dynamics of the unnormalized post-measurement state $\tilde{\rho}_{Q,\bm{\phi}_n}^{\text{diss}}(t)$ is characterized by a family of completely positive and trace non-increasing (CPTNI) maps. These maps are CP divisible when $d|G(t)|^2/dt\leq0$ , which coincides with the criterion for the monotonic decrease of the trace distance. Therefore, the trace distance is still a valid indicator of quantum (non-)Markovianity in the general sense of CPTNI maps. Similar discussions can also be found in Ref.~\cite{PhysRevA.103.032223}.

We now derive the divisible criterion for the dissipative model in terms of completely-positive and trace non-increasing maps. A dynamical map is usually considered as a collections of CPTP maps $\Lambda(t;0)$, where $\Lambda(t;0)$ is CP divisible if for all $t,\tau \geq 0 $ when $\Lambda(t+\tau;0)$ can be decomposed as 
\begin{equation}
\Lambda(t+\tau;0) =\Lambda(t+\tau;t)\Lambda(t;0), 
\end{equation}
where $\Lambda(t+\tau;t)$ is also a CPTP map. We relax the trace-preserving condition for this definition of CP divisibility because the dynamics for the dissipative model is described by CPTNI maps.
According to Eq.~\eqref{unnorm_post_measured_state_diss}, the Choi representation of map $\Lambda^{\text{diss}}_{\bm{\phi}_n}(t+\tau;t)$ can be derived as 
\begin{equation}
M^{\text{diss}}_{\bm{\phi}_n}(t+\tau;t)=
\begin{pmatrix}
1 & 0 & 0 & \frac{G(t+\tau)^*}{G(t)^*}\\
0 & \bar{N}\big[ 1 - \big|\frac{G(t+\tau)}{G(t)}\big|^2\big] &0 &0\\
0 & 0 & 0& 0
\\
\frac{G(t+\tau)}{G(t)} & 0 & 0 &\big|\frac{G(t+\tau)}{G(t)}\big|^2
\end{pmatrix},
\end{equation}
where $\bar{N} = \frac{N}{(N-2n)^2}$. Then, $\Lambda^{\text{diss}}_{\bm{\phi}_n}(t+\tau;t)$ is CP if and only if $|G(t+\tau)|^2\leq |G(t)|^2$. Therefore, $\Lambda^{\text{diss}}_{\bm{\phi}_n}(t+\tau;0)$ is CP divisible (in a sense of CPTNI maps) if and only if
\begin{equation}
\frac{d |G(t)|^2}{dt}\leq 0.
\end{equation}
\section{General expression for the overlap-integral expression\label{general_zeno}}

\begin{table*}
\begin{ruledtabular}
\begin{tabular}{lll}
  & Usual quantum Zeno effect & Space-time dual Zeno effect\\ [5pt]\hline
Setup & \vtop{\hbox{\strut A temporal sequence of measurements}\hbox{\strut  on one atom.}} &  \vtop{\hbox{\strut Only one measurement on $N$ paths}\hbox{\strut taken by the atom in an interferometer.}}\\ [15pt]\hline
Behavior of the filter functions&\vtop{\hbox{\strut The filter functions smear out}\hbox{\strut when increasing the measurement frequency.}} & \vtop{\hbox{\strut The filter functions remain localized}\hbox{\strut when increasing the number of paths.}}
\end{tabular}
\end{ruledtabular}
\caption{\label{zeno_st_zeno} Comparison of usual quantum Zeno effect and the space-time dual Zeno effect proposed in this work.}
\end{table*}

In the main text, we propose a space-time dual to quantum Zeno effect~(see in Table.~\ref{zeno_st_zeno}) and show that the corresponding decay factors for the dissipative and the pure dephasing models can be characterized by an overlap integral. We now provide a general expression of the decay factor. 

In general, the Hamiltonian for the qubit $Q$ and the environment $\mathcal{E}_i$ can be written as
\begin{align}
H_{Q\mathcal{E}_i} 
= H_Q + H_{\mathcal{E}_i} + H^I_{Q\mathcal{E}_i} \nonumber\\
\text{with}~~~~~H^I_{Q\mathcal{E}_i}=\sum_\alpha A_{i,\alpha}\otimes B_{i,\alpha}.
\end{align}
Here, $H_Q$ and $H_{\mathcal{E},i}$ respectively represent the free Hamiltonians for the qubit and the environment $\mathcal{E}_i$, and $H^I_{Q\mathcal{E}_i}$ denotes the interaction Hamiltonian, where $A_{i,\alpha} = A_{i,\alpha}^\dagger$ and $B_{i,\alpha} = B_{i,\alpha}^\dagger$ are the operators for $Q$ and $\mathcal{E}_i$, respectively. 

Within the interaction picture, the interaction Hamiltonian reads
\begin{align}
V^I_{Q\mathcal{E}_i}(t) &= \sum_\alpha A_{i,\alpha} (t) \otimes B_{i,\alpha}(t) \nonumber \\
\text{with}~~~A_{i,\alpha}(t)&=e^{i H_Q t}A_{i,\alpha} e^{-i H_Q t}, \nonumber \\
B_{i,\alpha}(t)&=e^{i H_{\mathcal{E}_i} t} B_{i,\alpha} e^{-i H_{\mathcal{E}_i} t}.
\end{align}
The propagator that governs the single-path time evolution can be written as
\begin{equation}
    U_{Q\mathcal{E}_i}(t) = \mathcal{T}_+ \exp\big[-i\int_0^t dt^\prime V^I_{Q\mathcal{E}_i}(t^\prime) \big].
\end{equation}
We now assume that the propagator can be approximated to the second-order perturbation such that 
\begin{equation}
U_{Q\mathcal{E}_i}(t) \approx \mathbb{1}+U_{Q\mathcal{E}_i,1}(t) + U_{Q\mathcal{E}_i,2}(t),
\end{equation}
where
\begin{align}
U_{Q\mathcal{E}_i,1}(t)&=-i\int_0^t dt_1 V^I_{Q\mathcal{E}_i}(t_1) \nonumber \\
\text{and}~~
U_{Q\mathcal{E}_i,2}(t) &= -\int_0^t dt_1 \int_0^{t_1} dt_2 V^I_{Q\mathcal{E}_i}(t_1)V^I_{Q\mathcal{E}_i}(t_2).
\end{align}

The time-dependent terms $\rho_{Q,i,j}(t)$ described in Eq.~\eqref{rho_Q_t_unnorm} can then be expanded to the second order, i.e., 
\begin{equation}
    \rho_{Q,i,j}(t) \approx \rho_Q(0) + \rho_{Q,i,j,1}(t)+\rho_{Q,i,j,2}(t).
\end{equation}
Here, $\rho_{Q,i,j,k}(t)$ characterizes the $k-$th order correction, wherein
\begin{widetext}
\begin{align}
    \rho_{Q,i,j,1}(t)&=\tr_{\{\mathcal{E}_k\}}\Big[U_{Q\mathcal{E}_i,1}(t)\rho_Q(0)\bigotimes_{l=1}^N\rho_{\mathcal{E}_l}(0) +\rho_Q(0)\bigotimes_{l=1}^N\rho_{\mathcal{E}_l}(0)U_{Q\mathcal{E}_j,1}^\dag(t) \Big] \nonumber \\
    &=\tr_{\mathcal{E}_i}[U_{Q\mathcal{E}_i,1}(t)\rho_{\mathcal{E}_i}(0)]\rho_Q(0)+ \rho_Q(0)\tr_{\mathcal{E}_j}[\rho_{\mathcal{E}_j}(0)U_{Q\mathcal{E}_j,1}^\dag(t)]\nonumber\\
    &=-i\sum_{\alpha,\beta}\int_0^t~dt_1\left\{A_{i,\alpha}(t_1)\rho_Q(0)\tr[B_{i,\alpha}(t_1)\rho_{\mathcal{E}_i}(0)]-\rho_Q(0)A_{j,\beta}(t_1)\tr[\rho_{\mathcal{E}_j}(0)B_{j,\beta}(t_1)]\right\},\nonumber\\
    \rho_{Q,i,j,2}(t)&=\tr_{\{\mathcal{E}_k\}}\Big[U_{Q\mathcal{E}_i,2}(t)\rho_Q(0)\bigotimes_{l=1}^N\rho_{\mathcal{E}_l}(0)+\rho_Q(0)\bigotimes_{l=1}^N\rho_{\mathcal{E}_l}(0)U_{Q\mathcal{E}_j,2}^\dagger(t)+U_{Q\mathcal{E}_i,1}(t)\rho_Q(0)\bigotimes_{l=1}^N\rho_{\mathcal{E}_l}(0)U_{Q\mathcal{E}_j,1}^\dagger(t)\Big]\nonumber\\
    &=-\sum_{\alpha,\beta}\int_0^t dt_1\int_0^{t_1}dt_2\Big\{A_{i,\alpha}(t_1)A_{j,\beta}(t_2)\rho_Q(0)\tr_{\{\mathcal{E}_k\}}\Big[B_{i,\alpha}(t_1)B_{j,\beta}(t_2)\bigotimes_{l=1}^N \rho_{\mathcal{E}_l}(0)\Big] \nonumber\\
    &~~~~~~~~~~~~~~~~~~~~~~~~~~~~~~+\rho_Q(0)A_{j,\beta}(t_2)A_{i,\alpha}(t_1)\tr_{\{\mathcal{E}_k\}}\Big[\bigotimes_{l=1}^N \rho_{\mathcal{E}_l}(0)B_{j,\beta}(t_2)B_{i,\alpha}(t_1)\Big]\Big\}\nonumber\\
    &~~~+\sum_{\alpha,\beta}\int_0^t dt_1 \int_0^t dt_2~ A_{i,\alpha}(t_1)\rho_Q(0)A_{j,\beta}(t_2)\tr_{\{\mathcal{E}_i\}}\left[B_{i,\alpha}(t_1)\bigotimes_{l=1}^N\rho_{\mathcal{E}_l}(0)B_{j,\beta}(t_2)\right].
\end{align}
\end{widetext}

For a large class of open-system models, the terms $\tr[B_{i,\alpha}(t)\rho_{\mathcal{E}_i}(0)]$ vanish, which implies $\rho_{Q,i,j,1}(t)=0$ and 

\begin{align}
    \rho_{Q,i,j,2}(t)=&\delta_{i,j}\sum_{\alpha,\beta}\int_0^t dt_1\int_0^{t_1} dt_2 \Big\{[A_{i,\beta}(t_2)\rho_Q(0)A_{i,\alpha}(t_1)\nonumber \\
    &~~~~~~-A_{i,\alpha}(t_1)A_{i,\beta}(t_2)]C_{i,\alpha,\beta}(t_1,t_2)+h.c.\Big\}.
\end{align}
Here, $h.c.$ denotes the Hermitian conjugate, and $C_{i,\alpha,\beta}(t_1,t_2)= \tr[B_{i,\alpha}(t_1) B_{i,\beta}(t_2)\rho_{\mathcal{E}_i}(0)]$ represents the two-point correlation function of the environment $\mathcal{E}_i$. Thus, we have
\begin{widetext}
    \begin{align}
        &\tilde{\rho}_{Q,\bm{\phi}}(t)=\frac{1}{N^2}\sum_{i,j}e^{-i(\phi_i-\phi_j)}\rho_{Q,i,j}(t)\nonumber\\
        &\approx\frac{1}{N^2}\left\{\sum_{i,j}e^{-i(\phi_i-\phi_j)}\rho_Q(0)+\sum_{i,\alpha,\beta}\int_0^t dt_1\int_0^{t_1}dt_2\left\{[A_{i,\beta}(t_2)\rho_Q(0)A_{i,\alpha}(t_1)-A_{i,\alpha}(t_1)A_{i,\beta}(t_2)\rho_Q(0)]C_{i,\alpha,\beta}(t_1,t_2)+h.c.\right\}\right\}.
    \end{align}
\end{widetext}

We assume that the corrections are sufficiently small, i.e., $|| \rho_{Q,i,j}(t) ||_\mathrm{tr}\ll1$, so that $\tr[\tilde{\rho}_{Q,\bm{\phi}}(t)] \approx \sum_{i,j}e^{-i(\phi_i-\phi_j)}/N^2$. Further, the correlation function usually takes the form $C_{i,\alpha,\beta}(t_1,t_2) = \int d\omega \mathcal{J}_i(\omega)f_{i,\alpha,\beta}(\omega,t_1,t_2)$, where $\mathcal{J}_i(\omega)$ denotes the coupling spectral density between the qubit and the environment $\mathcal{E}_i$, and $f_{i,\alpha,\beta}(\omega,t_1,t_2)$ summarizes the remaining information about the correlation function. Suppose that the initial qubit state is $\rho_Q(0)=|\psi\rangle \langle \psi|$. One can then express the decay factor in terms of averaging $N$ overlap integrals, namely,
\begin{align}
\frac{1}{N}\sum_{i=1}^N\int d\omega J_i(\omega) F_i(\omega,t,N,\bm{\phi}),
\end{align}
with the filter function
\begin{align}
&F_i(\omega,t,N,n)\nonumber\\
&=\frac{2N}{\sum_{k,l}e^{-i(\phi_k-\phi_l)}}\text{Re}\Big\{\sum_{\alpha,\beta}\int_0^t dt_1 \int_0^{t_1} dt_2~ f_{i,\alpha,\beta}(\omega,t_1,t_2)\nonumber\\
&~~~~~~~~~~~~~~~~~~~~~~~~~\tr\big[P_{\psi^\perp}A_{i,\beta}(t_2)\rho_Q(0)A_{i,\alpha}(t_1)\big] \Big\}.
\end{align}
Here, $P_{\psi^\perp}$ denotes a projector that projects onto the subspace orthogonal to $|\psi\rangle\langle \psi|$.

%

\end{document}